\documentclass[doublecol]{epl2}
\UseRawInputEncoding
\usepackage{amsmath}
\usepackage{amssymb}

\title{Probing the response of metals to low-frequency
{\boldmath$s$}-polarized evanescent fields}
\shorttitle{How to test the Drude response of metals to the $s$-polarized
low-frequency evanescent fields}

\author{G.~L.~Klimchitskaya\inst{1,2} \and V.~M.~Mostepanenko\inst{1,2,3}
 \and V.~B.~Svetovoy\inst{4}}
\shortauthor{G.~L.~Klimchitskaya \etal}

\institute{
  \inst{1}Central Astronomical Observatory at Pulkovo of the
Russian Academy of Sciences, Saint Petersburg,
196140, Russia\\
\inst{2}Peter the Great Saint Petersburg
Polytechnic University, Saint Petersburg, 195251, Russia\\
\inst{3}Kazan Federal University, Kazan, 420008, Russia\\
\inst{4}Frumkin Institute of Physical Chemistry and Electrochemistry,
Russian Academy
of Sciences, Leninsky prospect 31 bld. 4, 119071 Moscow, Russia
}

\abstract{
Experimental test for the response function of metals to the
low-frequency $s$-polarized evanescent waves is proposed
by measuring the lateral component of magnetic
field of an oscillating magnetic dipole spaced above a thick
metallic plate. This suggestion is motivated by the fact that
the Lifshitz theory using the Drude response function
is in contradiction with
high-precision measurements of the Casimir force performed at
separations exceeding 150~nm. Analytic expressions for the
lateral components of the magnetic field, which are fully
determined by the $s$-polarized evanescent waves, are reported
in terms of the reflection coefficients of the plate metal. Numerical
computations are performed for the reasonable values of the
experimental parameters for different models of the dielectric
response. The resulting fields differ by the orders of magnitude
depending on whether the Drude or plasma response function
is used in computations. Thus, the measurement of the magnetic
field will allow to discriminate between
these two options. Possible applications of the obtained results are
discussed.}

\begin{document}

\maketitle

\newcommand{\vk}{{\mbox{\boldmath$k$}}}
\newcommand{\rv}{{\mbox{\boldmath$r$}}}
\newcommand{\ve}{{\varepsilon}}
\newcommand{\okt}{{(\omega,k_t)}}

\section{Introduction}

The evanescent fields are created by the
oscillating currents and charges, which are always present in a
metal interior. It has been known that they are confined in the
vicinity of metallic surfaces and are characterized by the zero
mean Poynting vector. The respective solution to the wave equation
is characterized by at least one pure imaginary component of the
wave vector. This leads to the exponentially fast drop of the
evanescent wave in the corresponding spatial directions (see the
monographs \cite{1,2} for theory and diverse applications of
evanescent waves).

In spite of the paramount importance of some applications, e.g.,
as a tool to overcome the diffraction limit in optics \cite{3},
the electromagnetic response of metals to the $s$-polarized
evanescent waves is still not sufficiently investigated. A large
body of information is collected on surface plasmon
polaritons \cite{4} for large
magnitudes $k_t = |{\vk}_t|$ of the wave vector projection on
the surface. This information, however, refers solely to the
transverse magnetic ($p$-polarized) evanescent waves. At the same
time a much used technique of the total internal reflection and
frustrated total internal reflection enables to probe the
electromagnetic response of metals in the area of $k_t$ only
somewhat over the wave vector magnitude $k_0 = \omega/c$ where
$\omega$ is the field frequency \cite{5,6,7}. The reason is that
for transparent media the index of refraction in the infrared
domain is not sufficiently large (e.g., for Si it is equal to
only 3.4). Note also that the method of nano frustrated total
internal reflection used in the near-field optical microscopy
to overcome the diffraction limit \cite{3,8} is more sensitive
to the $p$-polarized evanescent waves \cite{9}.

The response of metals to the low-frequency electromagnetic
waves (including the evanescent ones) is routinely described
by means of the Drude model. Recently, however, the use of
this model in the case of $s$-polarized evanescent waves has
been questioned by a series of high-precision experiments on
measuring the Casimir force between metallic surfaces
performed at separations above 150~nm
\cite{10,11,12,13,14,15,16,17,18,19,20,21,22}. The point is
that the measurement data of these experiments are inconsistent
with theoretical predictions of the fundamental Lifshitz theory
\cite{23,24,25} if the electromagnetic response of metals at
low frequencies is described by the Drude model. The surprising
thing is that the same measurement data were found to be in
good agreement with theory if the plasma model is used although
this model disregards the relaxation properties of conduction
electrons and should not be applicable at low frequencies
\cite{10,11,12,13,14,15,16,17,18,19,20,21,22}. A more
sophisticated treatment of the problem revealed that the
major difference between the theoretical predictions for
the Casimir force using the Drude and plasma models
originates from the contribution of $s$-polarized
low-frequency evanescent waves \cite{26,27,28,29,30,31}
(see also reviews in \cite{32,33,34}). This places strong
emphasis on an independent test of the response function of
metals to the $s$-type evanescent waves.

In this Letter, we derive an analytic expression for the
magnetic field of an oscillating magnetic dipole spaced in
vacuum at some height above a thick metallic plate. The magnetic
moment of this dipole is pointed perpendicular to the plate.
According to our results obtained in the general case of a
spatially nonlocal electromagnetic response, the lateral
component of the magnetic field is determined by the
$s$-polarized low-frequency evanescent waves and is highly
sensitive to the character of response function of the metal.
Based on this, we propose an experimental test for
a response function to the
$s$-type evanescent waves. Computations are made for typical
experimental parameters. It is shown that the magnitudes of
lateral component of the dipole magnetic field computed using
the Drude and plasma models differ by up to a factor of several
thousands depending on the dipole oscillation frequency. This
allows to either confirm or exclude the theoretical description
of the response of metals to the $s$-polarized evanescent waves
by means of the Drude model. Possible implication of the
suggested test for the Casimir physics and in a wider context
for condensed matter physics and optics is discussed.

\section{Magnetic dipole above metallic plate}

First let us consider the oscillating magnetic dipole in vacuum in the
absence of metallic plate. We assume that it is at the origin of  the
coordinates $\rv_0=(0,0,0)$ and the magnetic moment
$\mbox{\boldmath$m$}=[0,0,m_0\exp(-i\omega t)]$ is directed along the
$z$-axis. On the assumption that the dipole size is much smaller than
the wavelength $\lambda=2\pi/\omega$, the components of its magnetic
field at a point $\rv=(x,y,z)\equiv(x_1,x_2,z)$  are given by \cite{35}
[here and below it is implied that all fileds depend on $t$ as
$\exp(-i\omega t)$]
\begin{eqnarray}
&&
H_{x_{\alpha}}(\omega,\rv)=-m_0\frac{x_{\alpha}z}{r^2}\left(
\frac{k_0^2}{r}+3i\frac{k_0}{r^2}-\frac{3}{r^3}\right)
e^{ik_0r},
\nonumber \\
&&
H_{z}(\omega,\rv)=m_0\left[
\frac{k_0^2}{r}+i\frac{k_0}{r^2}-\frac{1}{r^3} \right.
\label{eq1}\\
&&~~~~~~~~~~~~~~~~~\left.
-\frac{z^2}{r^2}\left(
\frac{k_0^2}{r}+3i\frac{k_0}{r^2}-\frac{3}{r^3}\right)\right]
e^{ik_0r},
\nonumber
\end{eqnarray}
\noindent
where $\alpha=1,\,2$, $k_0\equiv\omega/c$, $r=|\rv|=(x^2+y^2+z^2)^{1/2}$\!\!,
and the Gaussian system of units is used.

The components of the electric field of magnetic dipole in free space
are \cite{35}
\begin{equation}
\mbox{\boldmath$E$}(\omega,\rv)=im_0k_0\left(i\frac{k_0}{r^2}-\frac{1}{r^3}
\right)e^{ik_0r}\left(\begin{array}{rrr}y\\-x\\0\end{array}\right).
\label{eq2}
\end{equation}
\noindent
Note that the magnitudes of the electric field components are suppressed by
a factor $k_0r$ in comparison with those for the magnetic field.
 Since in the suggested experimental test $k_0r < 10^{-9}$ (see below), the
electric field of the dipole can be ignored.

In what follows we shall use the Fourier expansion of the field (\ref{eq1})
in two-dimensional plane waves
\begin{equation}
\mbox{\boldmath$H$}(\omega,\rv)=\frac{1}{(2\pi)^2}\int d\vk_t
e^{i\vk_t\rv_t} \mbox{\boldmath$H$}(\omega,\vk_t,z),
\label{eq3}
\end{equation}
\noindent
where $\vk_t=(k_x,k_y)\equiv(k_1,k_2)$ and $\rv_t=(x,y)\equiv(x_1,x_2)$.
The explicit expressions for the Fourier transform
$\mbox{\boldmath$H$}(\omega,\vk_t,z)$ can be easily found in the polar
coordinates using the integral representation for the Bessel functions
\cite{36} and several integrals \cite{37,38}. The results are
\begin{eqnarray}
&&
H_{x_{\alpha}}(\omega,\vk_t,z)=-2\pi i m_0k_{\alpha}{\rm sign}(z)
e^{-q|z|},
\nonumber \\
&&
H_{z}(\omega,\vk_t,z)=2\pi m_0\frac{k_{t}^2}{q}
e^{-q|z|},
\label{eq4}
\end{eqnarray}
\noindent
where $k_t^2=k_x^2+k_y^2$ and $q=(k_t^2-k_0^2)^{1/2}$\!\!\!.

\begin{figure}
\onefigure{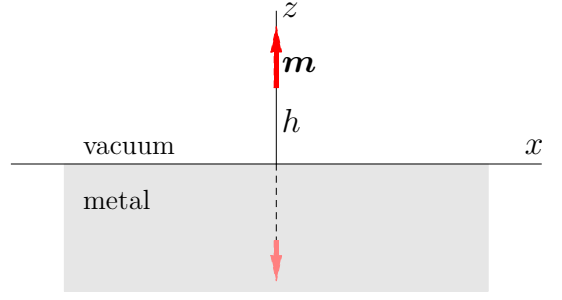}
\caption{Configuration of magnetic dipole spaced in vacuum above thick metallic
plate and the fictitious (image) dipole.}
\label{fig.1}
\end{figure}

We are coming now to the case of magnetic dipole located at the point
$(0,0,h)$ above the surface of a thick metallic plate coinciding with the
coordinate plane $z=0$ (see fig.~\ref{fig.1}).
The field of this dipole in the region above the
plate can be calculated by the method of images \cite{39,40}.
The same results are obtainable by the method of Green functions used
in the derivation of the Lifshitz formula for the Casimir force in \cite{23,24,25}.
According to the method of images, the field in question can be found as
a superposition of the fields of the real dipole and the fictitious
(image) dipole located at the point $(0,0,-h)$. In so doing the magnetic
moment of the image dipole depends on the reflectivity properties of metallic plate.

The reflectivity properties are usually described by the Fresnel reflection
coefficients defined for $p$- and $s$-polarizations of the electromagnetic field
\begin{equation}
R_p\okt=\frac{\ve(\omega)q-q_{\ve}}{\ve(\omega)q+q_{\ve}},
\quad
R_s\okt=\frac{q-q_{\ve}}{q+q_{\ve}},
\label{eq5}
\end{equation}
\noindent
where $q_{\ve}=[k_t^2-\ve(\omega)k_0^2]^{1/2}$ and $\ve(\omega)$ is the
frequency-dependent dielectric permittivity of the plate metal.
In this section, however, we do not use a specific form of the reflection
coefficients making the results applicable to both the local and nonlocal
response of the metal (note that the spatially nonlocal models
are the subject of considerable discussion in relation to the problems of
the Lifshitz theory mentioned above \cite{41,42,43,44,45,46,47,48,49,50,51,52}).
It has to be stressed also that in the configuration of fig.~\ref{fig.1}
the reflected wave is determined by the coefficient $R_s$ alone
because the electric field remains negligibly small.

Now we apply eq.~(\ref{eq4}) to both the dipole and its image in order to find the
Fourier transform of the magnetic field above the surface of metallic plate.
After taking into account that in the coordinate system of fig.~\ref{fig.1} one
has to replace $z$ with $z-(-h)=z+h$ for the fictitious dipole and with $z-h$ for
the real one, the result is
\begin{eqnarray}
&&
H_{x_{\alpha}}^{(p)}(\omega,\vk_t,z)=-2\pi i m_0k_{\alpha}\left[
R_s\okt e^{-q(z+h)}\right.
\nonumber \\
&&~~~~~~~~~~~~~~~~~~~~~~~~~\left.
+{\rm sign}(z-h)
e^{-q|z-h|}\right],
\label{eq6} \\
&&
H_{z}^{(p)}(\omega,\vk_t,z)=2\pi m_0\frac{k_{t}^2}{q}
\left[ R_s\okt e^{-q(z+h)}\right.
\nonumber \\
&&~~~~~~~~~~~~~~~~~~~~~~~~~~~~~~~~~\left.
+e^{-q|z-h|}\right].
\nonumber
\end{eqnarray}
\noindent

Substituting eq.~(\ref{eq6}) into eq.~(\ref{eq3}) and using
eq.~(\ref{eq1}) for the second
terms on the right-hand side of eq.~(\ref{eq6}), we find the magnetic field in the
domain above the metallic plate
\begin{eqnarray}
&&
H_{x_{\alpha}}^{(p)}(\omega,\rv)=
-\frac{im_0}{2\pi}\int d\vk_t R_s\okt k_{\alpha}
e^{i\vk_t\rv_t-q(z+h)}
\nonumber \\
&&~~~~~~~
-m_0\frac{x_{\alpha}(z-h)}{r^2}\left(
\frac{k_0^2}{r}+3i\frac{k_0}{r^2}-\frac{3}{r^3}\right)
e^{ik_0r},
\nonumber \\
&&
H_{z}^{p)}(\omega,\rv)=\frac{m_0}{2\pi}\int d\vk_t R_s\okt \frac{k_t^2}{q}
e^{i\vk_t\rv_t-q(z+h)}
\label{eq7} \\
&&~~~~~~~~~~
+m_0\left[
\frac{k_0^2}{r}+i\frac{k_0}{r^2}-\frac{1}{r^3}\right.
\nonumber \\
&&~~~~~~~~~~~~~~\left.
-\frac{(z-h)^2}{r^2}\left(
\frac{k_0^2}{r}+3i\frac{k_0}{r^2}-\frac{3}{r^3}\right)\right]
e^{ik_0r},
\nonumber
\end{eqnarray}
\noindent
where now $r=[x^2+y^2+(z-h)^2]^{1/2}\!\!\!$.

Introducing the polar coordinates under the integrals on the right-hand sides
 in eq.~(\ref{eq7}) and integrating over the angle variable \cite{36,37},
one arrives at
\begin{equation}
\hspace*{-5.5mm}
H_{x_{\alpha}}^{(p)}(\omega,\rv)=
\frac{m_0x_{\alpha}}{r_t}\!\!\int_0^{\infty}\!\!\! dk_tk_t^2
J_1(k_tr_t)R_s\okt e^{-q(z+h)}
\nonumber
\end{equation}
\begin{eqnarray}
&&~~~
-m_0\frac{x_{\alpha}(z-h)}{r^2}\left(
\frac{k_0^2}{r}+3i\frac{k_0}{r^2}-\frac{3}{r^3}\right)
e^{ik_0r},
\label{eq8} \\
&&
\hspace*{-4mm}
H_{z}^{p)}(\omega,\rv)=m_0\int_0^{\infty}\!\! dk_t\frac{k_t^3}{q}
J_0(k_tr_t)R_s\okt e^{-q(z+h)}
\nonumber \\
&&~~~~~~~~~~
+m_0\left[
\frac{k_0^2}{r}+i\frac{k_0}{r^2}-\frac{1}{r^3}\right.
\nonumber\\
&&~~~~~~~~~~~~~~
-\left.\frac{(z-h)^2}{r^2}\left(
\frac{k_0^2}{r}+3i\frac{k_0}{r^2}-\frac{3}{r^3}\right)\right]
e^{ik_0r},
\nonumber
\end{eqnarray}
\noindent
where $J_n$ is the Bessel function and $r_t=(x_1^2+x_2^2)^{1/2}$ is the polar
radial variable.

In the strict sense, both the propagating waves (for them $k_t<k_0$ and $q$
is pure imaginary) and the evanescent waves (for them $k_t>k_0$ and $q$ is
real) contribute to the field (\ref{eq8}). For the propagating waves,
however, the power in the exponential factors under the integral is pure
imaginary: $-q(z+h)=i(k_0^2-k_t^2)^{1/2}(z+h)$. Since in the suggested
experimental test it holds $k_0r\ll 1$ or, equivalently, $r\ll\lambda$,
it is possible to neglect by the phase in these factors. As a result,
one can see that the contribution of propagating waves to the integrals
in eq.~(\ref{eq8}) is smaller than that of the evanescent ones by the factor
of $1/(k_0h)^3=\lambda^3/(2\pi h)^3$. Below we choose $h\sim r$ in the
suggested experimental test leading to  $1/(k_0h)^3\sim 10^{27}$.
Because of this one can neglect by the contribution of propagating waves
and replace the lower integration limits in  eq.~(\ref{eq8}) with $k_0$.

Another important observation is that the contribution of the
real dipole to the total magnetic field above the plate (\ref{eq8})
vanishes for the lateral components $H_{x_{\alpha}}^{(p)}$ and becomes
simpler for the $z$-component $H_{z}^{(p)}$ if the calculation is
performed at the dipole height $z=h$. Since we are looking for a quantity
which is the most sensitive to the response function of metal to the
evanescent waves, there is no better candidates than the lateral
components of magnetic field given by
\begin{equation}
H_{x_{\alpha}}^{(p)}(\omega,\rv)=
\frac{m_0x_{\alpha}}{r_t}\!\!\int_{k_0}^{\infty}\!\!\!  dk_tk_t^2
J_1(k_tr_t)R_s\okt e^{-2qh}.
\label{eq9}
\end{equation}

In fact eq.~(\ref{eq9}) is analogous to the Lifshitz formula for the Casimir force
at large separations.
In this case, the difference in Casimir forces calculated using the Drude and
plasma dielectric functions is completely determined by different contributions
of the $s$-polarized evanescent waves. Thus, calculating
the field components (\ref{eq9})
for different models of the dielectric response of metal and
comparing the obtained results
with the measurement data, taken at $z=h$, it is possible to independently test
the validity of these models in the range of low-frequency evanescent waves.

\section{Parameters of magnetic dipole}

The magnetic dipole considered in the previous section can be realized in the
form of an alternating current $I_0\exp(-i\omega t)$ which flows through a
small circular loop of radius $R$ or a coil containing $N$ loops. In such
a situation, the magnitude of the magnetic moment is given by
\begin{equation}
m_0=\frac{1}{c}\pi NI_0R^2.
\label{eq10}
\end{equation}
\noindent
It is assumed that the coil size is much smaller than the height $h$ above
the plate where it is situated (see fig.~\ref{fig.1}).

The qualitative assessment for the desirable parameters of the coil can
be made under a simplified assumption of an ideal metal plate, i.e., by putting
$R_s\okt=-1$. In this case from eq.~(\ref{eq6}) we obtain the Fourier transform
of the $x$-component of magnetic field calculated at  height $z=h$ above
the plate
\begin{equation}
H_x^{(p)}(\omega,\vk_t,h)=2\pi im_0k_xe^{-2hq}.
\label{eq11}
\end{equation}

By comparing this with eq.~(\ref{eq4}), one can conclude that under a condition
$k_0r\ll 1$ the $x$-component of magnetic field is given by the last term on the
right-hand side of eq.~(\ref{eq1}) with $z=2h$
\begin{equation}
H_x^{(p)}(x)=\frac{6m_0xh}{(x^2+4h^2)^{5/2}}=\frac{m_0}{h^3}
\frac{6\tilde{x}}{(4+\tilde{x}^2)^{5/2}},
\label{eq12}
\end{equation}
\noindent
where we introduced the dimensionless variable $\tilde{x}=x/h$.

{}From (\ref{eq12}) it is seen that in order to maximize the lateral
projection of magnetic field one should have larger $m_0$ (i.e., higher
current and larger loop radius and the number of loops) but smaller
distance between the dipole and the plate.

It is rather difficult to create the magnetic dipole which combines a high
electric current with small $h$ because the coil size is supposed to be
much smaller than $h$. Fabrication of a small coil with a high current
and large number of loops presents a real challenge.
Micro electromagnets which meet our requirements have been, however, created
using the methods of micro \cite{53,54,55} and mini \cite{39} technologies.
The typical micro electromagnet may be of 1~mm height and consist of $N =10$
loops of $R=1~$mm radius. The coil of this kind is able to support the
current up to $I_0=3\times10^9~$stat\,A which is equal to 1~A in the SI.
Calculating the respective magnetic moment by (\ref{eq10}), one finds
\begin{equation}
m_0=3.14\times 10^{-2}~\mbox{erg/Oe}=3.14\times 10^{-5}~\mbox{A\,m}^2.
\label{eq13}
\end{equation}

To estimate the lateral magnetic field produced by this magnetic moment,
we substitute eq.~(\ref{eq13}) into eq.~(\ref{eq12}) and choose $x=h=10~$mm which
leads to $\tilde{x}=1$. The result is
\begin{equation}
H_x^{(p)}(x)=3.36\,\mbox{mOe}=3.36\times 10^{-7}\,\mbox{T}
=0.27\,\mbox{A\,m}^{-1}\!.
\label{eq14}
\end{equation}

Now we determine the desirable values of the oscillation frequency of the
magnetic dipole which depend  on the plate material. Good metals, such as
copper, are preferable for our purposes. The plasma frequency and the
relaxation parameter of copper are equal to \cite{57}
$\omega_p=1.12\times 10^{16}~$rad/s and $\gamma=1.38\times 10^{13}~$rad/s,
respectively. At low frequencies the dielectric permittivity is well
described by the Drude model
\begin{equation}
\ve_D(\omega)=1-\frac{\omega_p^2}{\omega(\omega+i\gamma)}.
\label{eq15}
\end{equation}

The oscillation frequency of the magnetic dipole has to satisfy two requirements.
First, it has to provide greater reflectance on the plate. Second, it has to give
large difference between the predicted field components when one uses
alternative models of the dielectric response of metal.

To comply with these requirements, we introduce the quantity
\begin{equation}
w=hq=h\sqrt{k_t^2-k_0^2}
\label{eq16}
\end{equation}
\noindent
and bring the reflection coefficient $R_s$ defined in eq.~(\ref{eq5}) to the form
\begin{equation}
R_s\okt=\frac{w-\sqrt{w^2-K(\omega)}}{w+\sqrt{w^2-K(\omega)}},
\label{eq17}
\end{equation}
\noindent
where
\begin{equation}
K(\omega)\equiv[\ve(\omega)-1]\frac{\omega^2}{\omega_h^2}
\label{eq18}
\end{equation}
\noindent
and $\omega_h=c/h$.

When the Drude model (\ref{eq15}) is used, the major contribution to
$|K(\omega)|$ at low frequencies is given by its imaginary part
\begin{equation}
 |K(\omega)|=\frac{\omega_p^2\gamma\omega}{(\omega^2+\gamma^2)\omega_h^2}
 \approx\frac{\omega_p^2\omega}{\gamma\omega_h^2}
\label{eq19}
\end{equation}
\noindent
With the proviso that $|K(\omega)|\gg 1$, the magnitude of the reflection
coefficient (\ref{eq17}) is close to that given by the plasma model which
disregards the relaxation properties of conduction electrons
\begin{equation}
\ve_{pl}(\omega)=1-\frac{\omega_p^2}{\omega^2}.
\label{eq20}
\end{equation}
\noindent
As discussed in Introduction, the Lifshitz theory of Casimir force comes to
agreement with the measurement data of high-precision experiments when the
plasma model is used as a response function of metals to the low-frequency
$s$-polarized evanescent waves.

If, to the contrary, $|K(\omega)|\ll 1$, the magnitude of the reflection
coefficient (\ref{eq17}) becomes too small and only a small share of the
dipole field is reflected from the plate. Because of this the dipole
oscillation frequency should satisfy the condition
\begin{equation}
 |K(\omega)|\approx\frac{\omega_p^2\omega}{\gamma\omega_h^2}
 \leqslant 1, \quad
 \omega\leqslant\Omega\equiv\frac{\gamma\omega_h^2}{\omega_p^2}.
\label{eq21}
\end{equation}

For $h=10~$mm and the Drude parameters of copper listed above, one finds
$\Omega\approx 100~$rad/s. Based on this, it is reasonable to consider the
dipole oscillation frequencies equal to 2, 10, and 100~rad/s, which lead to the
measurable magnetic fields.

\section{Computational results}

We have performed numerical computations of the lateral component of the
field $H_x^{(p)}$ for a magnetic dipole with the magnetic moment (\ref{eq13})
spaced above thick copper plate at the height of $h=10~$mm as shown in
fig.~\ref{fig.1}.
All computations were made by eq.~(\ref{eq9}) for $y=0$, i.e., $r_t=x$,
using the dielectric permittivities of the Drude model (\ref{eq15}), plasma model
(\ref{eq20}), and the spatially nonlocal phenomenological dielectric permittivity
introduced \cite{52} to reach an agreement between the Lifshitz theory and the
measurement data of high-precision experiments
do not disregarding the relaxation properties of conduction
electrons. We recall that eq.~(\ref{eq9}) describes the field component at the
height $h$ above the plate at a distance $x$ from the magnetic dipole.

\begin{figure}
\onefigure{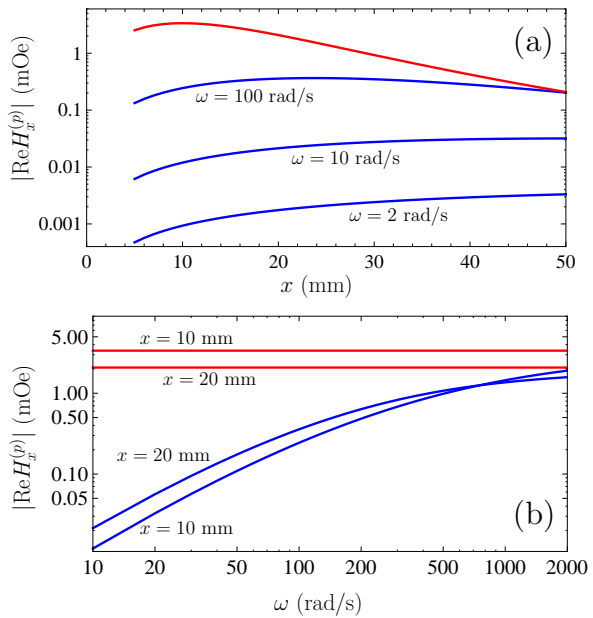}
\caption{The magnitude of real part
of the $x$-component of magnetic field of the
magnetic dipole reflected from the copper plate is shown (a) as a
function of separation by the three curves counted from bottom computed
using the Drude model for different dipole oscillation frequencies and
by the top curve computed using the plasma model and (b) as a function
of frequency by the bottom and top pairs of curves computed using the
Drude and plasma models, respectively, at two different separations.
}
\label{fig.2}
\end{figure}

The computational results for the magnitude of real part of $H_x^{(p)}$ are shown in
fig.~\ref{fig.2}(a) as a function of separation from the magnetic dipole.
The three curves counted from the bottom are computed
using the Drude model (\ref{eq15}) for the dipole oscillation frequencies
equal to 2, 10, and 100~rad/s, respectively. The top curve, which does not
depend on frequency, is computed by means of the plasma model (\ref{eq20}).
In fig.~\ref{fig.2}(b), the magnitude of the real part of $H_x^{(p)}$ is shown
as a function of the dipole oscillation frequency. The bottom pair of curves is
computed using the Drude model at 10 and 20~mm separation from the magnetic
dipole whereas the top pair of lines is computed at the same separations
be means of the plasma model. In both figs.~\ref{fig.2}(a) and \ref{fig.2}(b),
the computational results obtained using the spatially nonlocal model \cite{52}
are indistinguishable from those shown for the plasma model.

{}From fig.~\ref{fig.2} it is seen that the computational results essentially
depend on whether the Drude or the plasma model is used in computations.
Thus, for $|{\rm Re}H_x^{(p)}|$ the respective difference may be up to a factor
of $10^3$ depending on the dipole oscillation frequency. With increasing frequency,
this difference decreases. In order to deal with not-too-small fields, we
consider the oscillation frequency of 100~rad/s. In this case in the interval
from 10 to 20~mm one obtains
\begin{equation}
|{\rm Re}H_x^{(p)}|\lesssim 0.336\,\mbox{mOe}=3.36\times 10^{-8}\,\mbox{T}
=0.027\,\mbox{A\,m}^{-1}\!
\label{eq22}
\end{equation}
\noindent
if the Drude model is used in computations. If, however, the plasma model is used,
the values of $|{\rm Re}H_x^{(p)}|$ at the same separations are by more than an
order of magnitude larger. In doing so the parameter $k_0r$ discussed above is of the
order of $10^{-9}$. This allows clear experimental discrimination between the two
models of dielectric response to the low-frequency $s$-polarized evanescent waves
(note that the current limit for the resolution of weak magnetic fields is down
to $10^{-13}~$T \cite{58,59,60}).

Figure \ref{fig.2}(b) also demonstrates a profound effect of the used dielectric
model on the value of $|{\rm Re}H_x^{(p)}|$. Thus, the largest deviation between
the theoretical predictions obtained using the Drude and plasma models by the
factor of 280 is obtained for $\omega=10~$rad/s at $x=10~$mm.
For $\omega=100~$rad/s these predictions still differ by the factor of 14.

\begin{figure}
\onefigure{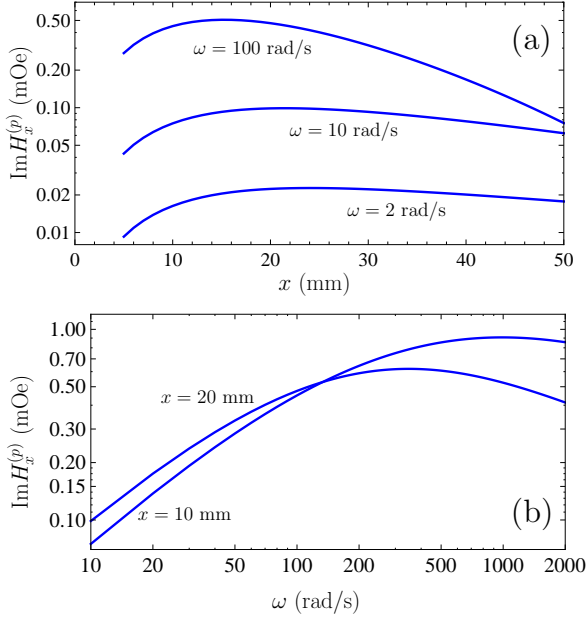}
\caption{The imaginary part
of the $x$-component of magnetic field for
copper plate is shown (a) as a
function of separation by the three curves computed
using the Drude model for different dipole oscillation frequencies
and (b) as a function
of frequency by two curves computed using the
Drude  model at two different separations. For the plasma
model, the imaginary part of the field is zero. }
\label{fig.3}
\end{figure}

Similar computations with all the same parameters using the Drude model have been
performed for ${\rm Im}H_x^{(p)}$. In fig.~\ref{fig.3}(a), the obtained results
are shown as a function of separation by the three curves from bottom to top for
the dipole oscillation frequency equal to 2, 10, and 100~rad/s, respectively.
In fig.~\ref{fig.3}(b), ${\rm Im}H_x^{(p)}$ is plotted as a function of frequency
by the two curves computed at the separation from the dipole of 10 and 20~mm.
If the plasma model is used, the reflection coefficient $R_s\okt$ is real
resulting in ${\rm Im}H_x^{(p)}=0$ [for the spatially nonlocal model \cite{52}
the values of ${\rm Im}H_x^{(p)}$ are by the factor of $10^{-11}$ smaller than
those shown by the bottom line in fig.~\ref{fig.3}(a)].

{}From fig.~\ref{fig.3} it is seen that if computations are performed by means
of the Drude model, the obtained values of ${\rm Im}H_x^{(p)}$ are much greater
than of $|{\rm Re}H_x^{(p)}|$ shown in fig.~\ref{fig.2} for all values of the
oscillation frequency. It is evident also that any nonzero measured value of
${\rm Im}H_x^{(p)}$ could be used in support of the Drude model and for an
exclusion of the plasma model as a response function to the low-frequency
$s$-polarized evanescent waves.

\section{Conclusions}

In the foregoing, it was argued that there
are no sufficient experimental evidences in favor of the statement
that the response of metals to the low-frequency $s$-polarized
evanescent waves is described by the dielectric permittivity of
the Drude model. An additional importance to this statement is
added by the fact that the Lifshitz theory of the Casimir force
becomes inconsistent with the measurement data of a number of
high-precision experiments if the electromagnetic response of
metals at low frequencies is described by the Drude model.
An agreement between the experiment and theory is restored if,
instead, the plasma model is used. Thus, a direct experimental
test of the response function of metals to the $s$-polarized
evanescent waves is crucial for both the Casimir effect
and for all numerous applications of these waves in fundamental
physics and in modern technology.

According to the obtained results, the direct experimental test
for the Drude dielectric permittivity as a response function
of metals to the low-frequency $s$-polarized evanescent waves
can be performed by measuring the lateral component of
magnetic field of the oscillating magnetic dipole spaced
above a thick metallic plate. To prove this statement, we have
derived an analytic expression for the magnetic field of a
small magnetic dipole located in the vicinity of metallic
surface and demonstrated that its lateral component is
determined by the contribution of the $s$-polarized
evanescent waves alone. Numerical computations of the real
and imaginary parts of the lateral field component were
performed for a copper plate for the reasonable experimental
parameters by describing the electromagnetic response of
metals by the Drude or plasma models for different oscillation
frequencies of the dipole. It was shown that the obtained
results differ by the orders of magnitude depending on the
model of dielectric permittivity used. Thus, by measuring the
field components, whose values are well within the limits of
experimental sensitivity, one could either validate or disprove
the possibility to use the Drude model in the range of
low-frequency $s$-polarized evanescent waves.

The results of the proposed experiment would be elucidating
for the long-standing problems of the Lifshitz theory and will
find applications in the fluorescence microscopy, infrared
spectroscopy, theory of superlenses, surface plasmon polaritons
and other prospective subjects of optics and condensed matter
physics.

\acknowledgments
 G.~L.~K.\ and
V.~M.~M.\ were partially supported by the Peter the Great Saint
Petersburg Polytechnic University in the framework of the Russian state
assignment for basic research (Project No.\ FSEG-2020-0024). The work
of V.~M.~M.\ was also supported by the Kazan Federal University
Strategic Academic Leadership Program. V.~B.~S.~was partially supported
by the Russian Science Foundation, grant No.\ 20-19-00214.


\end{document}